\input harvmac
\input labeldefs.tmp
\writedefs
\overfullrule=0pt

\input epsf
\def\fig#1#2#3{
\xdef#1{\the\figno}
\writedef{#1\leftbracket \the\figno}
\nobreak
\par\begingroup\parindent=0pt\leftskip=1cm\rightskip=1cm\parindent=0pt
\baselineskip=11pt
\midinsert
\centerline{#3}
\vskip 12pt
{\bf Fig.\ \the\figno:} #2\par
\endinsert\endgroup\par
\goodbreak
\global\advance\figno by1
}
\newwrite\tfile\global\newcount\tabno \global\tabno=1
\def\tab#1#2#3{
\xdef#1{\the\tabno}
\writedef{#1\leftbracket \the\tabno}
\nobreak
\par\begingroup\parindent=0pt\leftskip=1cm\rightskip=1cm\parindent=0pt
\baselineskip=11pt
\midinsert
\centerline{#3}
\vskip 12pt
{\bf Tab.\ \the\tabno:} #2\par
\endinsert\endgroup\par
\goodbreak
\global\advance\tabno by1
}

\font\cmss=cmss10 \font\cmsss=cmss10 at 7pt
\def\R{\relax{\rm I\kern-.18em R}}
\def\Z{\relax\ifmmode\mathchoice
{\hbox{\cmss Z\kern-.4em Z}}{\hbox{\cmss Z\kern-.4em Z}}
{\lower.9pt\hbox{\cmsss Z\kern-.4em Z}}
{\lower1.2pt\hbox{\cmsss Z\kern-.4em Z}}\else{\cmss Z\kern-.4em Z}\fi}
\def\bra#1{\big< {#1} \big|\,}
\def\ket#1{\,\big| {#1} \big>}

\def\frac#1#2{{#1\over#2}}
\def\T{{\bf T}}
\def\V{{\bf V}}
\def\H{{\bf H}}
%
%
%
\lref\Fur{H.~Furstenberg, {\it Trans.~Am.~Math.~Soc.}~68, 377 (1963);
G.~Benettin, L.~Galgani, A.~Giorgilli and J.-M.~Strelcyn,
{\it Meccanica} 15, 9 (1980).}
\lref\Motzkin{T.~Motzkin, {\it Bull.~Amer.~Math.~Soc.}~54, 352 (1948).}
\lref\SA{D.~Stauffer and A.~Aharony, {\it Introduction to percolation theory}
     (Taylor \& Francis, 1992).}
\lref\FK{P.~W.~Kasteleyn and C.~M.~Fortuin,
     {\it J.~Phys.~Soc.~Jap.}~46 (suppl.), 11 (1969).}
\lref\BAX{R.~J.~Baxter, {\sl Exactly solved models in statistical mechanics}
      (Academic Press, London, 1982).}
\lref\NIE{B.~Nienhuis, in {\sl Phase transitions and critical phenomena}
 vol.~11, ed.~C.~Domb and J.~L.~Lebowitz (Academic, London, 1987).}
\lref\NIJS{M.~P.~M.~den Nijs, {\it J.~Phys.}~A 12, 1857 (1979).}
\lref\NRS{B.~Nienhuis, E.~K.~Riedel and M.~Schick, {\it J.~Phys.}~A 13,
L189 (1980).}
\lref\BPZ{A.~A.~Belavin, A.~M.~Polyakov and A.~B.~Zamolodchikov,
      {\it Nucl.~Phys.}~B 241, 333 (1984).}
\lref\SD{B.~Duplantier and H.~Saleur, {\it Nucl.~Phys.}~B 290, 291 (1987).}
\lref\CAR{J.~L.~Cardy, {\it J.~Phys.}~A 25, L201 (1992).}
\lref\SMI{S.~Smirnov, {\it C.~R.~Acad.~Sci.~Paris}, to appear (2001).}
\lref\HB{S.~Havlin and A.~Bunde, {\sl Percolation II}, in
     {\sl Fractals and disordered systems}, ed.~A.~Bunde and S.~Havlin
     (Springer, Berlin, 1991).}
\lref\GRA{P.~Grassberger, {\it J.~Phys.}~A 25, 5475 (1992).}
\lref\GRAb{P.~Grassberger, {\it Physica}~A 262, 251 (1999),
cond-mat/9808095 v2.}
\lref\JZ{J.~L.~Jacobsen and P.~Zinn-Justin, work in progress.}
\lref\LSW{G.~F.~Lawler, O.~Schramm and W.~Werner, math.PR/0108211.}
\lref\CARb{J.~L.~Cardy, {\it J.~Phys.}~A 16, L355 (1983).}
\lref\ADA{M.~Aizenman, B.~Duplantier and A.~Aharony, cond-mat/9901018.}%
\lref\BN{H.~W.~J.~Bl\"{o}te and M.~P.~Nightingale,
         {\it Physica} A 112, 405 (1982).}
\lref\BNb{H.~W.~J.~Bl\"{o}te and B.~Nienhuis, {\it J.~Phys.}~A 22, 1415
 (1989).} 
\lref\JC{J.~L.~Jacobsen and J.~L.~Cardy, {\it Nucl.~Phys.}~B 515, 701
 (1998).} %
\lref\DJLP{Vl.~S.~Dotsenko, J.~L.~Jacobsen, M.-A.~Lewis and M.~Picco,
{\it Nucl.~Phys.}~B 546, 505 (1999).}
\lref\SDb{H.~Saleur and B.~Duplantier, {\it Phys.~Rev.~Lett.} 58, 2325
 (1987).} %
\lref\CON{A.~Coniglio, {\it J.~Phys.}~A 15, 3829 (1982).}
\lref\SAW{I.~G.~Enting, {\it J.~Phys.}~A 13, 3713 (1980)\semi
      B.~Derrida, {\it J.~Phys.}~A 14, L5 (1981).}
%
\Title{
\vbox{\baselineskip12pt\hbox{\tt cond-mat/0111374}}}
{{\vbox {
\vskip-10mm
\centerline{A Transfer Matrix for the}
\vskip2pt
\centerline{Backbone Exponent of Two-Dimensional Percolation}
}}}
\medskip
\centerline{Jesper Lykke Jacobsen {\it and} Paul Zinn-Justin}\medskip
\centerline{\sl Laboratoire de Physique Th\'eorique et Mod\`eles
 Statistiques} \centerline{\sl Universit\'e Paris-Sud, B\^atiment 100}
\centerline{\sl 91405 Orsay Cedex, France}
\vskip .2in
\noindent Rephrasing the backbone of
two-dimensional percolation as a monochromatic path crossing problem,
we investigate the latter by a transfer matrix approach. Conformal invariance
links the backbone dimension $D_{\rm b}$ to the highest eigenvalue of
the transfer matrix $\T$, and we obtain the result $D_{\rm b}=1.6431 \pm
 0.0006$. For a strip of width $L$, $\T$ is roughly of size $2^{3^L}$, but
we manage to reduce it to $\sim L!$. We find that the value of $D_{\rm b}$ is
stable with respect to inclusion of additional ``blobs'' tangent to the
 backbone in a finite number of points.
\Date{11/2001}

\newsec{Introduction}
The critical behavior of percolation has attracted considerable interest in
 the mathematical physics literature over the last decades. Whereas most
 practical applications (such as studying the efficiency of oil extraction
 from a porous soil, or the fractal geometry of a strike of lightning) take
 place in three spatial dimensions, analytical progress has largely been
 confined to two dimensions \SA. Although of geometric origin, percolation
 fits in the framework of critical phenomena, and in particular the concept
 of universality should apply. One therefore expects the specific choice of a
 discrete model (bond or site percolation) and of the lattice structure (e.g.
 square or triangular) to be of no relevance to the determination of the
 critical exponents.

A large part of the progress made is due to the identification with the
$q \to 1$ limit of the $q$-state Potts model \SA. A very fruitful idea has
been to treat the latter in terms of its random cluster formulation \FK,
and further in terms of the loops surrounding the clusters \BAX. Applying
Coulomb gas (and related) methods to the loop model led to a range of exact
results around 1980 \NIE. In particular, the correlation length exponent
$\nu = \frac43$ \CON\ and the magnetic exponent $x_h = \frac{5}{48}$
\refs{\NIJS,\NRS} (the codimension of which is the fractal dimension of the
percolating cluster, $D=2-x_h=\frac{91}{48}$) were computed.

The next major advance followed from the advent of conformal field theory
\BPZ, which provides an appealing correspondence between the $q$-state
Potts model (for particular values of $q$) and the so-called minimal models.
For instance, the exponents $x_k = \frac{1}{12}(k^2-1)$ with $k \ge 2$
\refs{\SDb,\ADA} describing the asymptotic decay of the probability
$P_k(r) \sim r^{-2 x_k}$ of having $k$ loop segments connecting two narrow
regions over a distance $r \gg 1$ \NIE\ were found to fit in the Kac
table of conformal dimensions \SD. Another remarkable result is the
celebrated Cardy formula \CAR\ expressing certain path-crossing probabilities
in terms of hypergeometric functions.

More recently, percolation has attracted the interest of probabilists. In a
groundbreaking publication, Smirnov has proved that the scaling limit of
site percolation on the triangular lattice exists and is described by the
stochastic Loewner evolution with parameter $\kappa = 6$ \SMI.
Consequently, most of the results referred to in the above have now been
rigorously proved.

Nevertheless, a certain class of exponents have continued to resist the
physicists' attempts over the years. These are most conveniently defined by
considering bond percolation inside a large square, of which we imagine two
opposing sides to be connected to superconducting plates
(see Fig.~\busbar). Each percolating
bond is stipulated to possess a fixed and finite conductivity, and an
electric voltage is applied across the plates. At the percolation threshold
$p=p_{\rm c}$, the part of the network that supports a non-zero current
is known as the {\it backbone}, and its fractal dimension $D_{\rm b}$
determines a critical exponent $\tilde{x}_2 = 2-D_{\rm b}$. Near $p_{\rm c}$,
the conductivity of the network scales as $(p-p_{\rm c})^t$, defining the
conductivity exponent $t$. The latter can be connected to the fractal
dimension of random walks constrained to the percolating cluster, or to its
backbone, via the Einstein relation \HB.

A number of conjectures for $\tilde{x}_2$ have been falsified as numerical
simulations have become increasingly accurate. The benchmark thus far is
the Monte Carlo method of Grassberger \GRA\ in which the conducting part of
the cluster is identified using a clever recursive algorithm. Large-scale
simulations yield the value $\tilde{x}_2 = 0.3568 \pm 0.0008$ \GRAb. The
exponent $\tilde{x}_2$ is actually a member of a family of so-called
monochromatic path-crossing exponents $\tilde{x}_k$ \ADA, with the magnetic
exponent fitting in as $x_h = \tilde{x}_1$. The higher exponents
$\tilde{x}_k$, $k \ge 3$ are all unknown.

In the present publication we provide a numerical estimate of $\tilde{x}_2$
using an algorithm which is entirely different from that of Grassberger.
Using the reformulation of $\tilde{x}_2$ as a path-crossing problem, we
relate it to the largest eigenvalue of a linear operator (actually a
transfer matrix) that builds all possible percolation clusters supporting
at least $k=2$ mutually non-intersecting paths. We work in the geometry of
semi-infinite strips of width $L$, with $L \le 9$.

Our approach is interesting in several respects. First, the reformulation as
an eigenvalue problem makes direct contact with the predictions of conformal
field theory \CARb.
That Grassberger's recursive algorithm defines a conformally invariant
observable is not a priori obvious, but the fact that the transformation to
a path-crossing problem involves a conformal transformation and that we
here obtain a consistent value of $\tilde{x}_2$ shows that this is indeed
the case. One would then further expect $\tilde{x}_2$ to be the conformal
dimension of a primary operator $\hat{O}_2$ in some (presently unknown)
conformal field theory of percolation. In particular, the conformal tower
of $\hat{O}_2$ should possess descendents whose conformal dimensions are
integer-spaced with respect to $\tilde{x}_2$. We have checked this prediction
by examining the scaling of the first few eigenvalues of our transfer matrix
with system size. We shall present evidence of a level two descendent with
conformal dimension $2.35 \pm 0.1$, whereas there does not appear to be
a descendent at level one.

Second, the generalization of our method to the case of
more ($k \ge 3$) paths, or to the Potts model with $q \neq 1$ states, are
immediate. Results for these cases will appear in a separate publication \JZ.
Third, from a technical point of view we have had to tackle the major
 obstacle of writing a transfer matrix in which some degrees of freedom (the
 percolation clusters) must be summed over, whereas others (the paths) act as
 constraints on the former but must not themselves be summed over. Fourth, we
 have devised an algorithm which is naturally parallelizable.

Like Grassberger \GRAb\ we find that the data for $\tilde{x}_2$ are hampered
by strong (presumably non-analytic) corrections to scaling. As a
consequence our final result
\eqn\est{
 \tilde{x}_2 = 0.3569\pm0.0006
}
confirms that of \GRAb, but unfortunately does not improve its precision.
On the other hand, we have devised some variants of our algorithm in which
the constraint of mutual avoidance of the two paths is relaxed, so that
they are allowed to touch in some configurations
at vertices but not to share an
edge. Physically this means that we measure the fractal dimension of the
backbone with some ``blobs'' that are tangent to it
included. The surprising result is that this relaxation
of the original definition does not alter the value of $\tilde{x}_2$.

The paper is laid out as follows. In Sec.~2 we review the
reasoning leading from the original formulation of the backbone dimension
to that of a path-crossing problem, and we restate the latter in a strip
geometry. The construction of the corresponding transfer matrix, and of
its associated state space, is described in Sec.~3. In
Sec.~4 we transcribe this as an algorithm and discuss its implementation.
The data is analyzed and extrapolated to the $L \to \infty$ limit in
Sec.~5. The appendix displays
some transfer matrices produced for small system size.

{\it Note added}: When this work was being completed we became aware of
the preprint of Lawler, Schramm and Werner \LSW\ in which
$\tilde{x}_1 = \frac{5}{48}$ is established on a rigorous basis, following
Smirnov \SMI. The authors also relate $\tilde{x}_2$ to a second-order
partial differential equation with specific boundary conditions, but fail
to provide an explicit solution of the latter. We thank John Cardy for
bringing this to our attention.

\newsec{Path-crossing probabilities}
Let us return to the formulation of the backbone problem given in the
introduction, namely in the so-called busbar geometry (see Fig.~\busbar).
The condition that a given point (site or bond, as the case may be)
on the spanning cluster belongs to the
backbone is that it can be connected to either of the superconducting
plates by means of two mutually non-intersecting paths.
\foot{Strictly speaking, this condition includes also points which are
being held exactly at zero current in a Wheatstone's bridge-like
arrangement. Since in the continuum limit the percolation cluster is
almost surely ``asymmetric'', such points are extremely rare. See also
the discussion below on the possibility of contact points for paths.}

\fig\busbar{Busbar geometry, here shown for the case of bond percolation on
 the square lattice. The backbone is indicated by fat
 edges.}{\epsfxsize=6cm\epsfbox{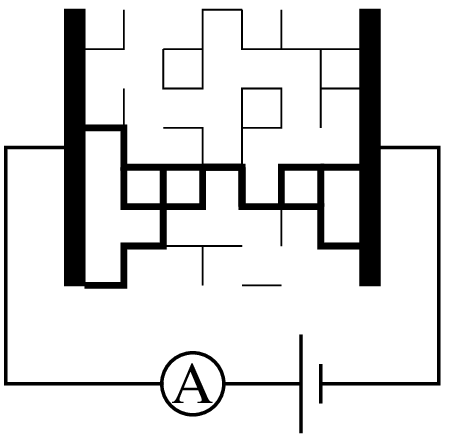}}

This choice of geometry is somewhat unnatural, as it does not fully display
the rotational symmetry of the continuum limit. It is more convenient to
work in an annular geometry limited by two concentric circles of radii
$r \ll 1$ and $R \gg 1$. Interpreting the inner circle as the point which
is a potential element of the backbone, and the
outer circle as the point at infinity, we see that a given percolating
configuration in the annulus contributes to the backbone if and only if the
two circles are connected by two mutually non-intersecting paths on the
percolating cluster(s); see Fig.~\annulus.

\fig\annulus{Annular geometry endowed with critical percolation (here shown
 in the continuum limit). A possible choice of two disjoint percolating paths
 is shown as dashed lines.}{\epsfxsize=6cm\epsfbox{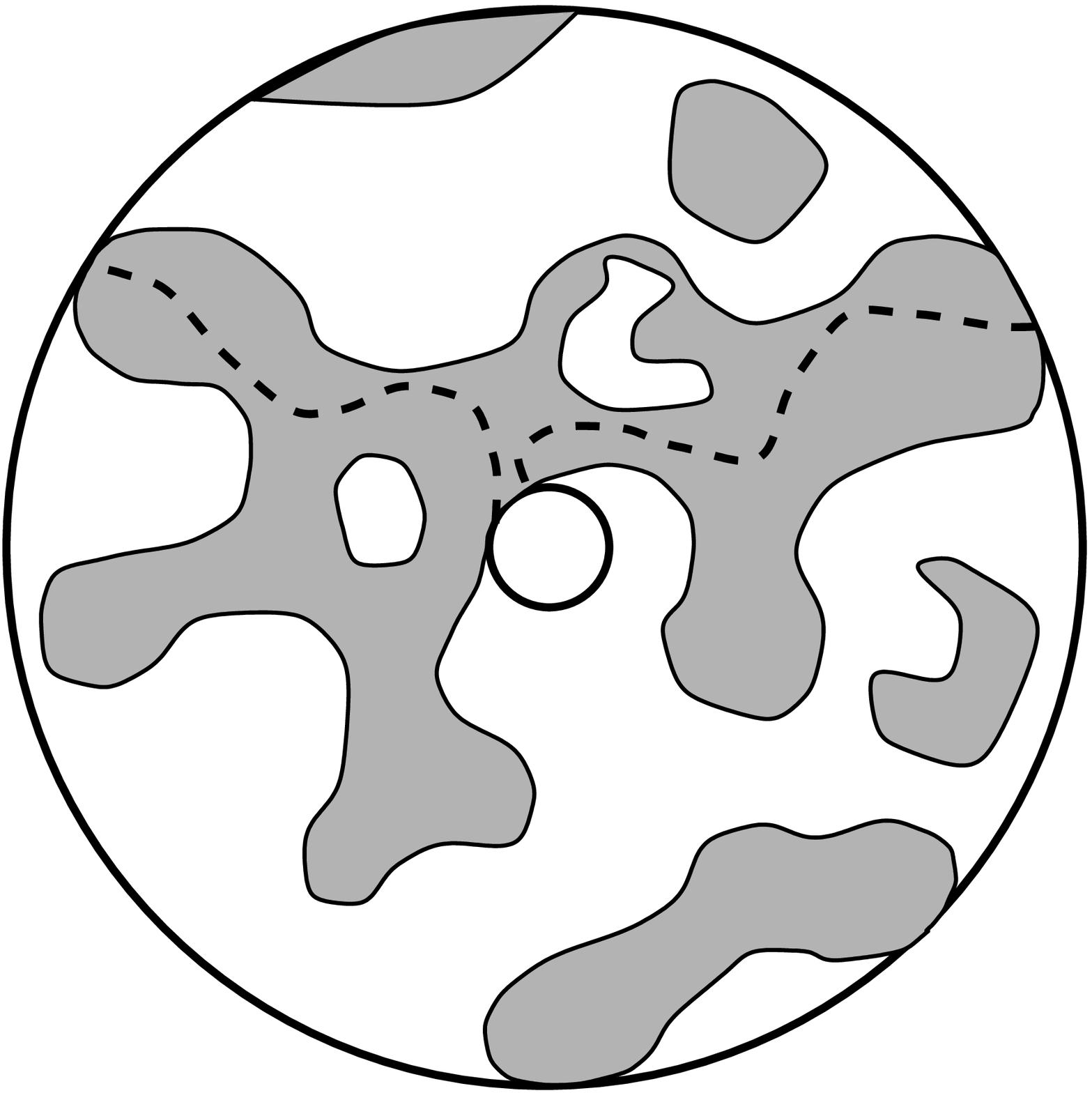}}

More generally, one may define higher exponents $\tilde{x}_k$ by studying,
at the percolation threshold,
the probability $P_k(r,R) \sim \left( \frac{r}{R} \right)^{\tilde{x}_k}$
that the annulus is traversed by $k$ mutually non-intersecting paths.
Clearly, the configurations in which these paths belong to different clusters
are asymptotically subdominant, and so we might as well assume that they
belong to the same cluster.

The situation may be further refined \ADA\ by considering path-crossing
 events in which a given number of traversing paths belong to the clusters,
 and the remaining number belong to the dual clusters.%
\foot{For site percolation, the dual clusters consist of the non-conducting
(uncolored, white) sites. For bond percolation on the square lattice, it is
most natural to think of the dual clusters in terms of the standard duality
transformation in the random cluster model \FK, according to which any
conducting edge is intersected by a non-conducting dual edge, and vice
 versa.} More precisely, for each $k$ there are $2^k$ types of path
 configurations, each specified by a set of color variables
 $(\tau_1,\tau_2,\ldots,\tau_k)$ with $\tau_i=+1$ (resp.~$\tau_i=-1$) meaning
 that path number $i$ belongs to the clusters (resp.~to the dual clusters).
 Within the context of the $q$-state Potts model (with $q \neq 1$), it is not
 obvious whether different choices of the color variables will lead to the
 same critical exponents, except of course for the obvious symmetries
 obtained by rotating the sequence $(\tau_1,\tau_2,\ldots,\tau_k)$, reversing
 its order, or dualizing it. But in the percolation case ($q=1$) the bonds
 (or sites) are uncorrelated, and various parts of the system may be dualized
 independently. Using
this approach, it has been proven in the case of site percolation on the
triangular lattice \ADA\ that all the polychromatic sequences (in which both
 a $\tau_i=+1$ and a $\tau_j=-1$ are represented; $k \ge 2$) share the same
 critical exponents. In particular, any polychromatic color configuration may
 be transformed into the alternating one, $\tau_i=(-1)^i$.

We expect this result to be independent of a particular lattice realisation,
and thus to apply also to bond percolation. In this case,
the identification of the critical exponent with that of $k$ traversing loop
segments on the surrounding lattice, referred to as $x_k$ in the
 introduction, becomes evident (at least for $k$ even). A rigorous proof that
 the formula $x_k=\frac{1}{12}(k^2-1)$ applies to the polychromatic path
 crossing problem for site percolation on the triangular lattice was spelled
 out in \ADA.

For monochromatic sequences (all $\tau_i=+1$) the argument given in \ADA\
fails, and the corresponding exponents $\tilde{x}_k$ are expected to be
different from the $x_k$. Indeed, from entropic considerations it should be
clear that $x_k < \tilde{x}_k < x_{2k}$.

Several of the $x_k$ have nice physical interpretations. Thus, $x_2$, $x_3$
and $x_4$ are respectively the codimensions of the cluster perimeter (hull)
\SDb, of the external (accessible) perimeter \ADA, and of the set of pivotal
(singly connecting) bonds \SDb. The latter also yields the correlation length
exponent \CON, via the scaling relation $\nu=1/(2-x_4)$.

In the absense of an exact solution, one might imagine evaluating the
exponents $\tilde{x}_k$ numerically by measuring the decay of the path
crossing probabilities on an annulus, as outlined above. A more feasible
alternative is to compute certain restricted free energies on semi-infinite
cylinders by using a transfer matrix, as we shall describe in the next
section. These free energies can be related to the critical exponents as
follows.

Since the scaling limit of critical percolation is conformally invariant
\refs{\BPZ,\SMI}, one is allowed to transform the annular geometry of
Fig.~\annulus\ into a cylindrical one by means of the conformal mapping $w
\equiv u+iv = \frac{L}{2\pi} \log(z)$. The transformed complex coordinate $w$
may be thought of as imbedded in the strip $-\infty<u<\infty$, $0 \le v \le
 L$ with periodic boundary conditions in the $v$-direction. All this means is
 that Fig.~\annulus\ must be viewed in perspective, interpreting the inner
 and outer circles as the extremities of the cylinder.

We are going to make use of the following result:
let $f_0(L)$ be the free energy per unit area for the
unrestricted percolation problem, and $\tilde{f}_k(L)$ (resp.~$f_k(L)$)
the corresponding quantity for the constrained problem where only those
configurations are included in the partition sum in which (at least)
$k$ monochromatic (resp.~polychromatic) paths span the length of a
semi-infinite cylinder of width $L$. Then as $L\to\infty$, the discrete
lattice model, at criticality, should have a continuum limit described
by conformal field theory, so that \CARb
\eqna\dimop
$$
\eqalignno{
 \tilde{f}_k(L) - f_0(L) &= \frac{2 \pi \tilde{x}_k}{L^2} + o(L^{-2}),
&\dimop{a}\cr
 f_k(L) - f_0(L)         &= \frac{2 \pi x_k}{L^2}         + o(L^{-2}).
&\dimop{b}\cr}
$$

We shall obtain
estimates for the $\tilde{x}_k$ by extrapolating data for sufficiently large
strips to the limit $L\to\infty$.

\newsec{Transfer matrix algorithm}
It has been known for a long time how to numerically compute the $f_k(L)$,
by writing the transfer matrix for the loop model \BAX\ in the basis of
planar (Catalan-like) connectivities (see \BNb\ for a closely
related computation). The same is true for $\tilde{f}_1(L)$ by using the
trick of adding a ghost site \BN, or alternatively (via a duality
argument) by forbidding the clusters to wrap around the cylinder \JC.

The computation of $\tilde{f}_2(L)$, the principle of which we now
describe, is considerably more complicated. The main complication stems
from the fact that to compute the corresponding partition sum we must
exclude those configurations of the percolation clusters that do not
support (at least) two spanning paths, and count each of those that do
with {\it unit weight} (and not with a weight equal to the number of
ways two such paths can be realized for the given cluster configuration).
Roughly speaking, the degrees of freedom are the clusters {\it and} the
paths, and we must trace over the former but {\it not} the latter.

For the sake of definiteness we consider in this section critical {\it bond}
percolation on a square lattice, though the principle of the transfer matrix
can be applied to any lattice with any probability of occupation $p$, and to
bond as well as site percolation. Since in our case $p_{\rm c}={1 \over 2}$
\SA, it is convenient to simply assign a weight of one to every
configuration of percolating/non-percolating bonds. For now we consider
the simplest orientation of the lattice, which corresponds to $L$ sites
in the transverse direction with periodic boundary conditions (Fig.~\latt).
With all these conventions, $f_0(L)=-2 \log 2$ in Eq.~\dimop{}.
We shall discuss later another possible orientation of the lattice.
\fig\latt{The square lattice with periodic boundary conditions along
one of its orientations. The dotted lines are time slices.}{
\epsfbox{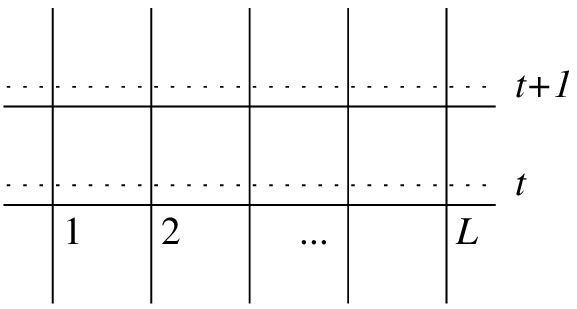}}

We keep track of the paths by defining path configurations
in analogy with
those used in the transfer matrix calculations of the self-avoiding walk
\SAW: among the $L$ sites in a row (at time $t=t_0$), two sites are
connected to the point at infinity (time $t=-\infty$) by means of paths.
Furthermore, in order to allow subsequent backtracking of either path
(at a later instant $t>t_0$), the remaining sites may be connected in
pairs by means of backward arches. The possible configurations for $L=4$
are listed on Fig.~\pathconfigs.
\fig\pathconfigs{The 10 possible path configurations for
 $L=4$.}{\epsfxsize=10cm \epsfbox{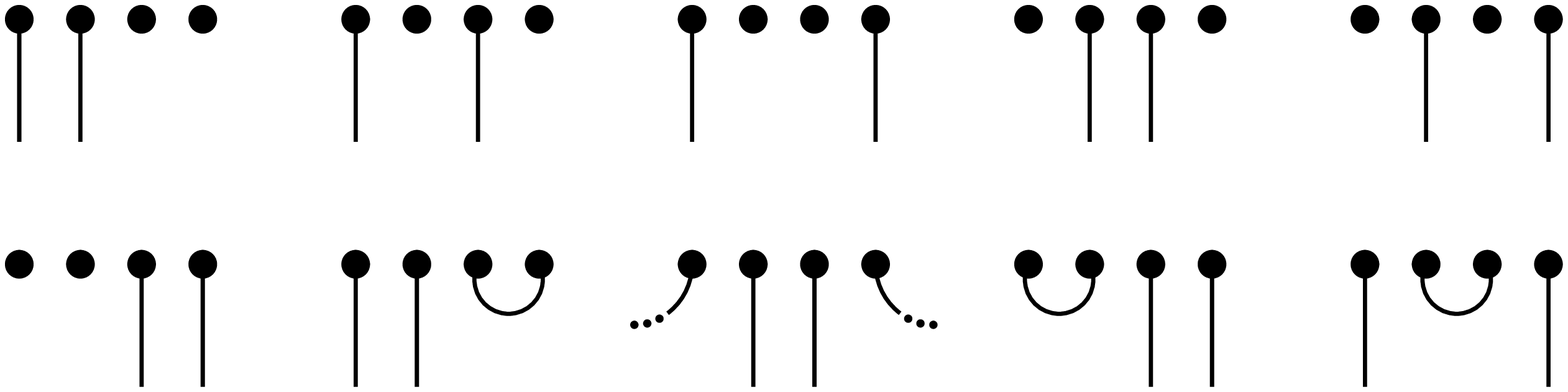}}

To overcome the difficulty of not summing over the possible path
configurations, we define the basis states on which the transfer matrix acts
as {\it lists} of path configurations. Elements of the list give all possible
realizations of the positions of the paths (and of the arches) which are
compatible with the ``past'' of the state.

Formally, if ${\cal P}$ is the set of path configurations,
then basis states are indexed by non-empty subsets of ${\cal P}$ (one
must exclude the empty subset since it corresponds to states
for which there is no possibility of two disjoint paths reaching time $t$).
Note that the
dimension of the total space is $2^{\# {\cal P}}-1$, which grows
extremely rapidly with $L$. We shall return to this point when we discuss
practical implementation.

By definition, the matrix element $\T_{{\cal A}{\cal B}}$ between
basis states indexed by ${\cal A}\subset {\cal P}$
and ${\cal B}\subset {\cal P}$ equals the number of
configurations of the bonds between time $t$ and time $t+1$,
such that the state ${\cal A}$ at time $t+1$ is obtained from the state
 ${\cal B}$ at time $t$. Given the initial state ${\cal B}$ and the
 configuration of the bonds $\omega\in\Omega$
(that is whether they are percolating or not, $\Omega$ being the
set of all possibilities),
the procedure to determine the final state ${\cal A}$
is as follows:
\item{$\star$} For each possible initial path configuration $b\in {\cal B}$,
consider all possible continuations of the existing lines at time $t$
(the two original
paths and the arches) that are compatible with the configuration of the bonds
$\omega$. Note that each line must be either continued to a site at time
 $t+1$, or be connected to another line (in which case it will reemerge at
 the other end of the arch; the lines coming from infinity or from the
same arch cannot be connected to each other).
Furthermore, for each pair of adjacent
empty sites one must consider the possibility of creating a new arch.
Let $\phi(b,\omega)\subset {\cal P}$ be the list of path configurations
at $t+1$ thus produced.
\item{$\star$} The full state ${\cal A}$ is reconstructed
by simply putting together
all the possibilities (of the form $\phi(b,\omega)$, $b\in{\cal B}$)
obtained for each initial path configuration.
If one finds ${\cal A}=\emptyset$, this means that no continuation
is possible, and the state is excluded.

We give an example of such a computation on Fig.~\evol.
\fig\evol{Evolution of two path configurations with the
same percolation configuration. Solid (resp.\ dashed) lines
represent percolating (resp.\ non-percolating) bonds,
whereas thick lines represent the possible paths.}{\epsfxsize=12cm
\epsfbox{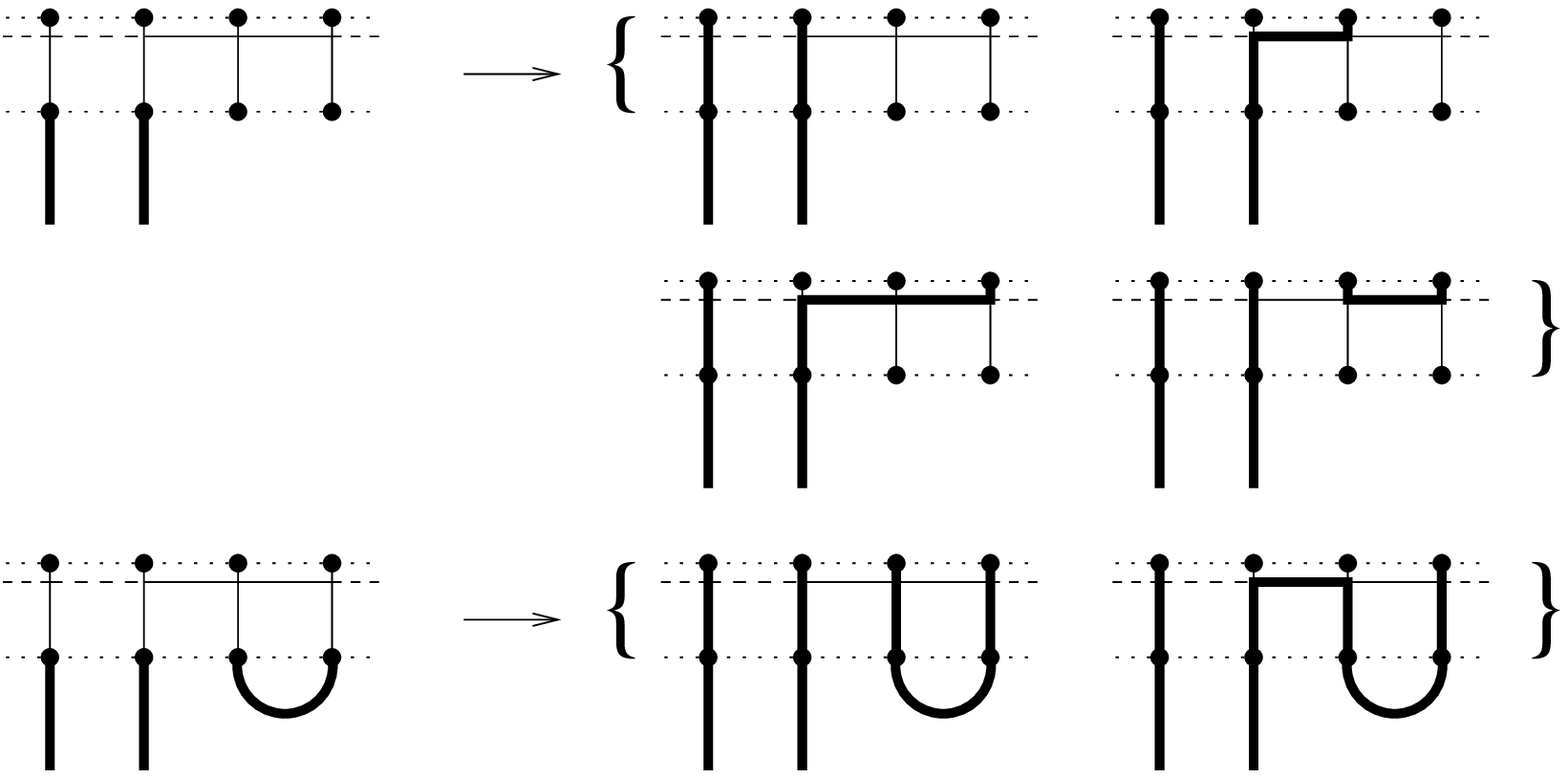}}

In other words, we have the formal identity:
$$\T\ket{\cal B}=\sum_{\omega\in\Omega}\ket{\bigcup_{b\in {\cal B}}
 \phi(b,\omega)}$$ which shows quite explicitly that one sums over bond
 configurations but not over path configurations.

Finally, the free energy per unit area is given by
\eqn\conn{
\tilde{f}_2(L)=-\lim_{t\to\infty} {1\over Lt} \log \bra{\cal
 A}\T(L)^t\ket{\cal B} }
where the states ${\cal A}$, ${\cal B}$ specify the boundary conditions and
are essentially arbitrary (the state ${\cal A}$ should belong to the image
of $\T$, see next section),
and $\T(L)$ is the transfer matrix for strip width $L$.

As $t\to\infty$, the matrix element $\bra{\cal A}\T(L)^t\ket{\cal B}$ is
dominated by the largest eigenvalue $\lambda(L)$ of $\T(L)$, and combining
Eqs.~\dimop{a} and \conn, we find:
\eqn\asy{
{1\over2^{2L}}\lambda(L)=1-{2\pi\tilde{x}_2\over L} + o(L^{-1})
}

\newsec{Algorithmic details}
In order to appreciate how effective the transfer matrix approch is,
it is important to understand the structure of the matrix constructed in
the previous section. It is an integer-valued matrix of extremely large size,
 but many of its entries are zero. In fact, starting from any basis state
 $\ket{\cal B}$, a very
limited number of states are generated. These are the only states that matter
for the determination of the largest eigenvalue(s) and we can thus
restrict ourselves to a submatrix of much smaller size.

We now describe schematically the procedure we used. The main steps
of the algorithm are as follows:

\item{(i)} Start with an arbitrary basis state (ideally, one that we know is
generated by iteration of the transfer matrix).
Put it onto a ``stack'' of states to process.
\item{(ii)} Pick a state ${\cal B}$ from the stack and ``process'' it,
i.e.\ generate the non-zero entries $\T_{\cal AB}$, and store them.
This encodes one column of the transfer matrix.
\item{(iii)} Consider every new basis state ${\cal A}$ that has been
 generated at step (ii); check if it has already been processed; if not, add
 it to the stack. If the stack is non-empty, go back to step (ii).
\item{(iv)} Finally, once the stack is empty, the largest
eigenvalue is computed by simple iteration of the matrix that has been
generated.

The transfer matrix is such that the submatrix thus generated
has no zero rows or columns. We call this submatrix the reduced transfer
 matrix.

An important remark for practical applications is that this procedure
is highly parallelizable: several CPUs can perform step (ii) simultaneously
and independently, only the stack must be shared. In practice,
it is necessary to have a server that communicates with the various clients
involved in the computation; it ensures that their stacks are synchronized,
and dispatches the tasks.
At the end of each calculation (step (ii)), a client sends the server
the new states created and receives the states created by other clients
in the meantime. The time spent updating the stack being very small compared
to the calculation time, the parallelization is near 100\%\ efficient
(at least up to 20 clients which is the maximum we tested).

Let us now discuss in more detail this procedure.

First, we must define how to encode path configurations. A study of
Fig.~\pathconfigs\ shows that if exactly $k=2$ paths are connected to
$t=-\infty$, then they can be considered as an extra arch. This trick reduces
the number of configurations and slightly
simplifies the implementation (but cannot be extended to $k\ne 2$). We can
then move the point at infinity and redraw the configurations as standard
 arch configurations%
\foot{The number of $L$-point arch configurations equals $m_L-1$, where
$m_L$ are the Motzkin numbers \Motzkin\ (the empty configuration is
 excluded). The generating function
$M(x)\equiv \sum_{L=0}^\infty m_L x^L = (1-x-\sqrt{1-2x-3x^2})/2x^2$,
has a singularity in $x=x_{\rm c}={1 \over 3}$, showing that the number
of path configurations is $m_L \approx 3^L$ asymptotically.}%
, see Fig.~\arches.
\fig\arches{The configurations of Fig.~\pathconfigs\ redrawn as
arches.}{\epsfxsize=10cm
\epsfbox{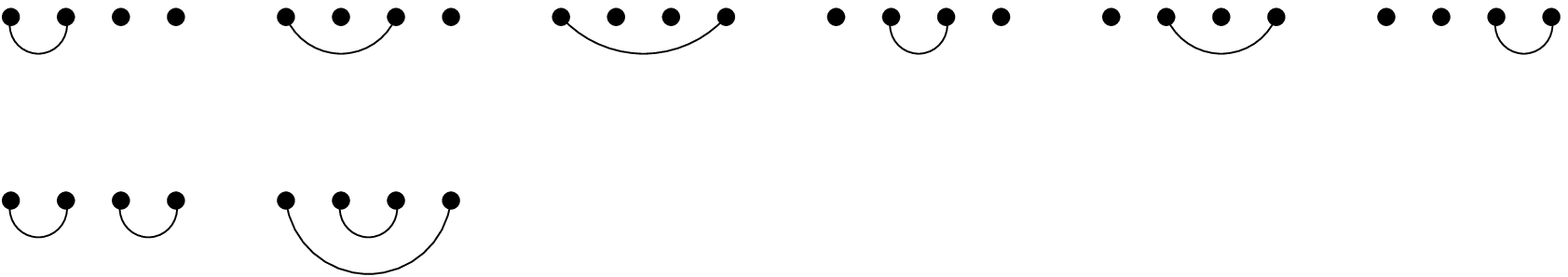}}
An arch configuration is then encoded in a standard way as a sequence of
closing / empty / opening steps, that is
$\epsilon_i\in \{ -1, 0, +1\}$, $1\le i\le L$,
such that the height function $h_\ell=\sum_{i=1}^\ell \epsilon_i$ satisfies
$h_\ell\ge 0$ for all $\ell$ and $h_L=0$. States are now defined as sorted
 lists of path configurations.

Next, we discuss how to perform step (ii) in practice. One possibility
would be to apply directly the principle of section 3, that is to consider
all possible bond configurations between 2 successive time slices and for
 each, to produce the resulting state. However, since there are $2^{2L}$ such
 configurations, the time required to do so grows exponentially, which is not
 satisfactory. Besides, the determination of all possible continuations of
 the paths to time $t+1$ is a rather complex task.
Instead, we shall use a factorization of the transfer matrix
as a product of $L$ sparse matrices $\T_i$, $1\le i\le L$ which
describe the addition of a single site.
The details of the factorization depend on the exact situation envisioned.
We present here three cases.

\subsec{The square lattice with standard orientation}
The example used so far is that of the square lattice with its
usual orientation. In this case the factorization
can be pushed further by writing that $\T(L)=\H_1\ldots \H_L\,\V_1\ldots
 \V_L$ where $\V_i$ (resp.\ $\H_i$) corresponds to the addition of a single
 vertical (resp.\ horizontal) bond (Fig.~\latttwo).
\fig\latttwo{Factorization of the transfer matrix.}{
\epsfbox{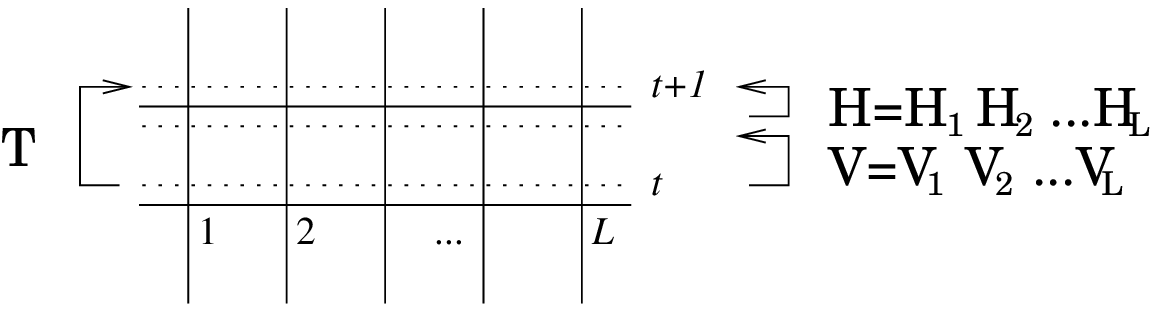}}

The action of $\V_i$ is very simple: $\V_i=\V_i'+\V_i''$ where
$\V_i'$ (resp.\ $\V_i''$)
describes the evolution when the vertical bond number $i$ is percolating
 (resp.\ non percolating). $\V'_i$ is simply the identity, whereas $\V_i''$
 acts
on path configurations as follows: either a path/arch is at site $i$,
in which case it gives $0$ (the path cannot cross the non-percolating
bond), or there is not and it is the identity. The action on a state
made of {\it several} path configurations can be deduced
from these basic rules, as explained in section 3.

The action of $\H_i$ is slightly more complicated: $\H_i=\H'_i+\H''_i$,
similarly as above. The definitions of $\H'_i$ and $\H''_i$ must take into
account all the possibilites of continuations, recombinations and creations
of paths along the horizontal bonds.
This requires working, as intermediate states, with path configurations
of length $L+2$ instead of $L$, since one must temporarily distinguish the
paths directed horizontally and vertically at the first and last vertices
being currently processed. We leave the details as an exercise to the
 interested reader.

A final ingredient is
that one can use the dihedral symmetry of the transfer matrix: since
the latter commutes with cyclic permutations of the sites and with
reflections, one can select a representative in each orbit of the dihedral
group among the basis states. Note that the action is an overall action
on all configurations that constitute the state simultaneously.
The states generated by the procedure above can then be replaced with
the representative state of their orbit, producing a smaller transfer matrix
but with identical eigenvalues. This further reduces the size of the transfer
matrix, by a factor of (roughly) $L$.

\subsec{The square lattice with standard orientation 2:
the square/octogon deformation}
It is interesting to study variants of the algorithm above. One natural
question is: if one allows the paths to touch each other at {\it vertices},
how is the asymptotic behavior of the free energy
modified and in particular is $\tilde{x}_2$
left unchanged? Another possible formulation of this question is to
consider a deformation of the lattice in which each vertex is replaced with
a small square, resulting in a square/octogon lattice (Fig.~\lattthree). The
 bonds of the small square are always percolating and allow paths that would
 have touched at a vertex to avoid each other.\foot{Note that a path
crossing a vertex of the original lattice can correspond to two different
 paths on the deformed lattice,
but since we do not sum over path realizations this is of
no consequence.}
\fig\lattthree{Deformation of the square lattice.}{
\epsfbox{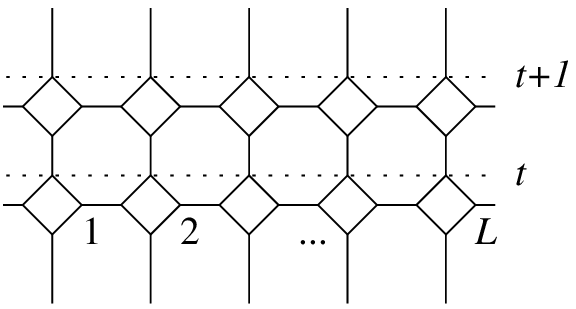}}

Physical insight suggests that such modifications
should not affect the values of $\tilde{x}_2$. The reason is that, just
like the Wheatstone bridges configurations mentioned in the introduction,
the fact that current flows through loops which are connected to
the backbone by just one point
is rather unstable since any microscopic defect that breaks the symmetry
between the two orientations of the loops (deforming the
lattice is precisely a way of introducing such a defect)
will produces a non-zero current.
If $\tilde{x}_2$ is to be universal it should not depend on such
microscopic details. It is this insight that we would like to test.
There is another, more practical reason one would want to study such
modifications of the algorithm, which will be apparent in section 5.

It is very simple to modify the transfer matrix of section 4.3 to allow such
path evolutions. $V_i$ is unchanged, whereas $H_i$ now allows
two paths to reach the same vertex and emerge from it
as if they had not touched each other.

\subsec{The square lattice with light-cone orientation:
the hexagon deformation}
Finally, we rotate the lattice by 45 degrees, the motivation being that
we expect better convergence properties, as observed empirically in similar
computations \refs{\JC,\DJLP}. Unfortunately, there is no efficient way
to encode the corresponding configurations, and we are therefore led
to a modification of the lattice which is similar to what was done in
section 4.2: this time the result is a hexagon lattice in which vertical
 bonds are always percolating (Fig.~\lattfour).
This is equivalent to allowing ``horizontal tangencies''
on the original square lattice, that is allowing two paths to touch
at one vertex in the configuration where the two upper edges belong to the
same path; however, ``vertical tangencies'' are still excluded.
\fig\lattfour{Another deformation of the square lattice.}{
\epsfbox{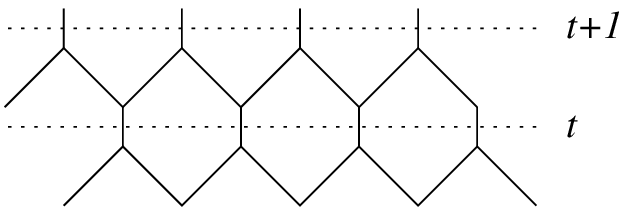}}
In this case, encoding the states becomes completely identical to what
was done previously. There is a decomposition $\T=\T_1\ldots \T_L$
where $\T_i$ adds an extra vertex $i$ at time $t+1$ (and two bonds).
Since the new sites at $t+1$ are now shifted with respect to the sites
at $t$, the action of the transfer matrix includes a conventional rotation
of a half-bond's length (or $\pi/L$).

Relations \conn--\asy\ must also be modified to take into account the 45
 degrees rotation; the latter introduces an extra factor of $2$ in the unit
 of area, so that $f_0(L)=-4\log 2$ and:
\eqn\asyb{
{1\over2^{2L}}\lambda(L)=1-{\pi\tilde{x}_2\over L} + o(L^{-1})
}
This factor of $2$ alone increases the accuracy of the measurement of
$\tilde{x}_2$ compared to the other two cases, since the corrections are
expected to be smaller.

\newsec{Numerical results}
We show on table \sizematters\ the size of the reduced
transfer matrix for $4\le L\le 9$, in the three cases presented above
(sections 4.1, 4.2, 4.3).
While the full matrix is very roughly of size $2^{3^L}$, the size of the
reduced matrix seems to grow as $L!$,
which is still large but not as intractable. It is interesting to note
that $s_2<s_1$, that is the modification of the lattice to allow
 configurations where paths touch at a point {\it decreases}\/ the number of
 states. \tab\sizematters{Size of the reduced transfer
 matrix.}{\vbox{\offinterlineskip
 \halign{\strut\hfil$#$\hfil\quad&\vrule#&&\quad\hfil$#$\hfil\crcr
L&&4&5&6&7&8&9\cr
\omit&height2pt\cr
\noalign{\hrule}
\omit&height2pt\cr
s_1&&15&72&515&4219&41728&?\cr
s_2&&12&51&291&1893&14923&132799\cr
s_3&&12&51&291&1893&14923&132799\cr
}}}
We have no deep explanation
for the remarkable equality of sizes of algorithms 2 and 3,
except the observed fact that the states generated are the same
in the two cases.

Next we present the data for the largest eigenvalue of the transfer matrix
on table \eigen\ with a twelve digit accuracy.
\def\squad{\hskip6pt}
\tab\eigen{Largest eigenvalue of the transfer matrix. The last row shows the
 second real eigenvalue for the third transfer
 matrix.}{\vbox{\offinterlineskip
 \halign{\strut\hfil$#$\hfil\squad&\vrule#&&\squad\hfil$#$\hfil\crcr
L&&4&5&6&7&8&9\cr
\omit&height2pt\cr
\noalign{\hrule}
\omit&height2pt\cr
\lambda_1/2^{2L}&&
\scriptstyle 0.514287790945&
\scriptstyle 0.594678112301&
\scriptstyle 0.653760363032&
\scriptstyle 0.698459489246&
\scriptstyle 0.733243927216&
\scriptstyle ?\cr
\lambda_2/2^{2L}&&
\scriptstyle 0.540388840500&
\scriptstyle 0.617254658842&
\scriptstyle 0.672285202673&
\scriptstyle 0.713573950794&
\scriptstyle 0.745682316102&
\scriptstyle 0.771356857232\cr
\lambda_3/2^{2L}&&
\scriptstyle 0.718747415570&
\scriptstyle 0.775012703547&
\scriptstyle 0.812529692986&
\scriptstyle 0.839330907375&
\scriptstyle 0.859432882632&
\scriptstyle 0.875067710677\cr
\lambda'_3/2^{2L}&&
\scriptstyle 0.058692638251&
\scriptstyle 0.145046191784&
\scriptstyle 0.224345992159&
\scriptstyle 0.292806902950&
\scriptstyle 0.351338353673&
\scriptstyle 0.40153182\cr
}}}
In order to study the asymptotic behavior of these series of numbers,
we use Eq.~\asy\ for cases 1 and 2 (or \asyb\ for case 3) to extract
approximate values of $\tilde{x}_2$. The results are on Fig.~\fits.
We also presented quadratic fits of these data.
\fig\fits{Values of $\tilde{x}_2$ obtained from the eigenvalues of
the transfer matrices (Tab.~\eigen). The order of colors is: 1 -- red,
2 -- green, 3 -- blue.}{\epsfxsize=6cm\epsfbox{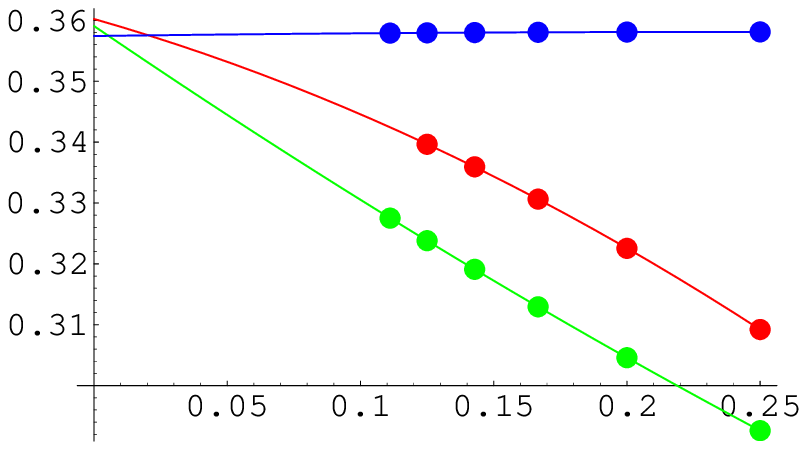}}

Several remarks are in order. First the two curves corresponding to the
 square lattice with its regular orientation (with or without contacts at
 points) seem to converge nicely within the range allowed by the fits. This
 means that the value of $\tilde{x}_2$ is {\it not}\/ affected by this
 modification. However, it is clear
that the next corrections to $\lambda_1$ and $\lambda_2$
are quite different. Secondly, it is again manifest on figure \fits\ that
the third set of data, corresponding to the 45 degrees rotated square
 lattice, reaches its limit much faster than the other two. Whereas various
 fits will give a limiting value for the first two anywhere between $0.355$
 and $0.36$, the range is limited to $0.3563$ to $0.3575$ for the latter.
 Assuming all these limits to be the same, we reach the estimate \est\
 mentioned in the introduction. Note that there is no simple way for us to
 evaluate error bars since the results are entirely dependent on the fits
 used, the latter being arbitrary without any knowledge about the subleading
 corrections.

Finally, numerical estimates of the norms of higher eigenvalues of the
transfer matrix spectra can be extracted by a standard
iteration/orthogonalization procedure \Fur. Using this method, complex
eigenvalues are characterized by an oscillatory behavior and can thus
be discarded (we expect physical observables to be linked to real
 eigenvalues). Specializing to case 3 (cf.~section 4.3 above), we find the
 fourth eigenvalue (in norm) to be the second real one. Its finite-size
 scaling is well fitted by \dimop{}, defining a critical index
\eqn\sublead{
\tilde{x}_2' = 2.35 \pm 0.1
}
This is consistent with the conformal dimension of a level two descendent of
the backbone operator.

Extracting the scaling dimensions for even higher eigenvalues becomes
increasingly problematic, as the finite-size effects get considerably
stronger. It should however be noticed that the third real eigenvalue
is doubly degenerate for any width $L \ge 4$. This is supposed to have
implications for the organization of the conformal tower of the backbone
operator.

\bigskip\centerline{\bf Acknowledgements}
J.L.J.\ is grateful to Jean Vannimenus for interesting discussions during 
an early stage of this project, and wishes to thank Hubert Saleur for
drawing his attention to the importance of studying higher eigenvalues.
P.Z.-J.\ would like to thank Claude Jacquemin for introducing him to
the mysteries of sockets. 

\appendix{A}{Structure of some small size transfer matrices}
As an illustration of the algorithm explained in this article, we provide
here the simplest non-trivial transfer matrices obtained
with the geometries of sections 4.2 and 4.3. They correspond to a strip
length $L=4$ and their size is $s=12$.

\fig\states{Basis states (up to overall dihedral transformations)
for $L=4$.}{\epsfxsize=15cm\epsfbox{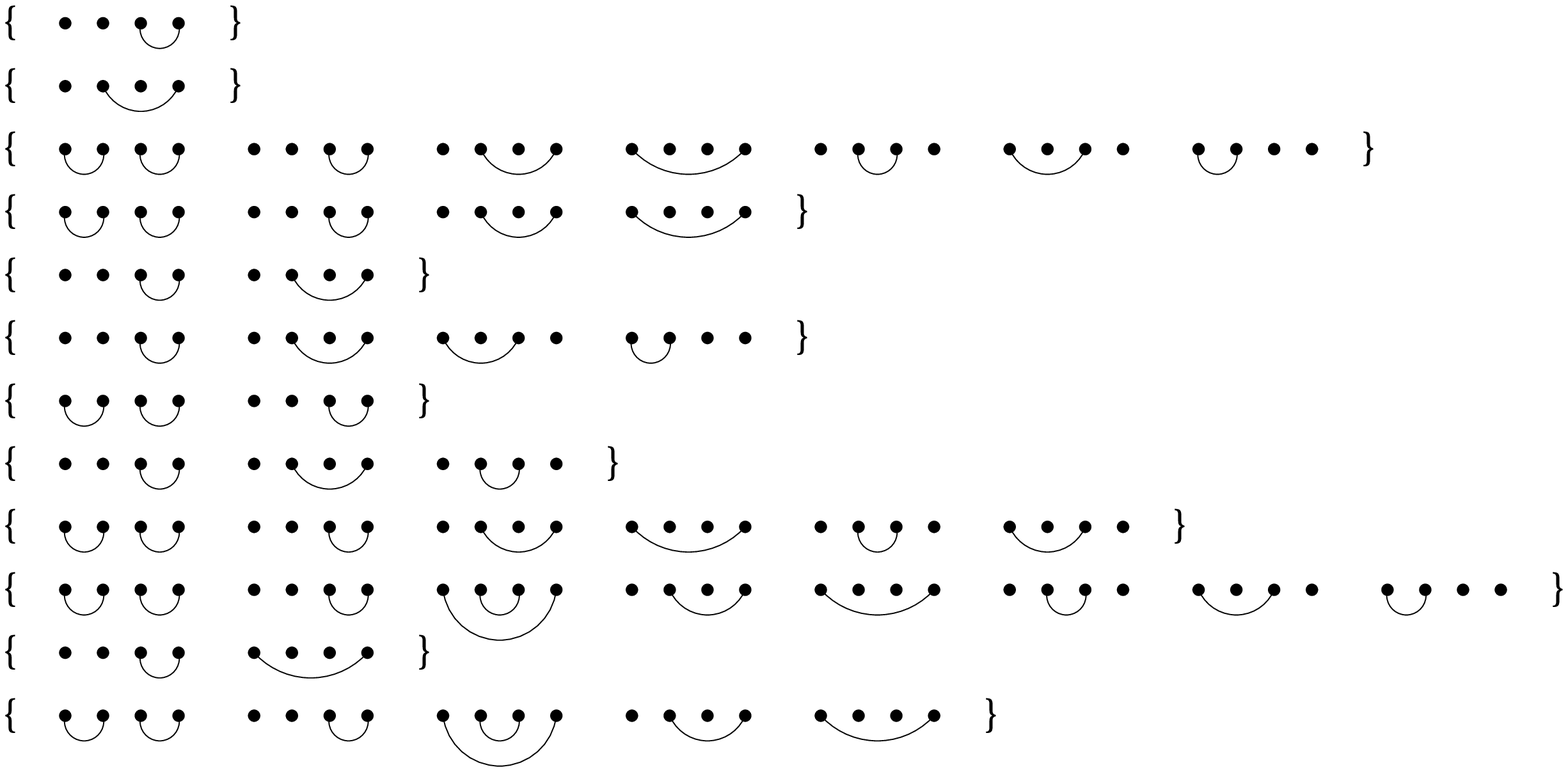}}
On Fig.~\states\ is described the {\it basis} in which these
matrices are expressed.

The matrices themselves read:
$$\T_2=\pmatrix{
8& 0& 8& 4& 4& 4& 6& 8& 6& 8& 8& 4\cr
0& 4& 2& 1& 2& 2& 0& 2& 2& 2& 0& 1\cr
4& 0& 8& 8& 4& 9& 5& 4& 5& 4& 8& 6\cr
8& 0& 8& 10& 6& 8& 10& 8& 8& 8& 12& 8\cr
8& 16& 16& 12& 16& 20& 6& 16& 18& 16& 8& 12\cr
4& 8& 8& 8& 10& 18& 3& 8& 11& 8& 4& 8\cr
8& 0& 8& 4& 4& 4& 10& 8& 6& 8& 8& 4\cr
8& 8& 28& 13& 12& 16& 6& 20& 22& 28& 14& 13\cr
8& 0& 24& 10& 6& 8& 10& 16& 20& 24& 12& 8\cr
8& 20& 62& 34& 26& 47& 8& 34& 56& 66& 18& 38\cr
0& 0& 0& 1& 0& 0& 0& 0& 0& 0& 2& 1\cr
0& 8& 4& 7& 6& 8& 0& 4& 6& 4& 2& 9\cr
}$$

$$\T_3=\pmatrix{
  36&   24&   32&   19&   28&   24&   24&   40&   28&   32&   33&   19\cr
   9&   18&   18&   12&   15&   16&    7&   18&   17&   18&   12&   12\cr
   2&    7&    8&   13&    8&   14&    6&    1&    2&    0&    9&    7\cr
  10&   12&    0&   14&   13&   12&   18&    6&    6&    0&   12&    8\cr
  36&   48&   24&   38&   49&   52&   30&   40&   38&   24&   34&   34\cr
   6&   11&    2&    9&   12&   18&    6&    5&    6&    2&    6&    9\cr
  12&    8&    0&    0&    6&    0&   12&    8&    0&    0&    7&    0\cr
  10&   12&   84&   39&   24&   44&   12&   48&   64&   84&   32&   44\cr
   6&    4&   32&   13&    9&   12&    6&   26&   30&   32&   14&   11\cr
   1&    0&   39&   16&    6&   19&    3&   16&   34&   47&    5&   27\cr
   0&    0&    0&    7&    3&    6&    4&    0&    3&    0&    6&    6\cr
   0&    0&    0&    7&    3&    6&    0&    0&    3&    0&    2&   10\cr}
$$

\listrefs
\bye